\documentclass{aa}
\usepackage{natbib}
\usepackage{graphicx}
\usepackage{txfonts}
\def\HH{\mbox{H$_2$}}
\def\fH2{\mbox{f$_{{\rm H}_2}$}}

\def\fDNM{\mbox{f$_{\rm DNM}$}}
\def\EBV{\mbox{E$_{\rm B-V}$}}
\def\AV{\mbox{A$_{\rm V}$}}

\def\nH2{\mbox{${\rm n}_\HH}$}
\def\pccc{~{\rm cm}^{-3}} 
 
\def\pcc{~{\rm cm}^{-2}}
\def\Tsub#1 {\mbox{${\rm T}_{\rm #1}$}}
\def\TK  {\Tsub K }

\def\Snu{\mbox{S$_{\nu}$}}
\def\arcsec{\mbox{$^{\prime\prime}$}} \def\arcmin{\mbox{$^{\prime}$}}

\def\degr{$^{\rm o}$}
\def\p{\mbox{$^+$}}

\def\cch{\mbox{C$_2$H}}

\def\h13cop{\mbox{{H$^{13}$CO\p}}}
\def\c3h2{\mbox{C$_3$H$_2$}}
\def\cc3h2{\mbox{{\it c}-C$_3$H$_2$}}
 
\def\G0{\mbox{G$_0$}}

\def\ddeg{{}^\circ\kern-.1em}

\def \kms{\mbox{km\,s$^{-1}$}}

\def\E#1 {$10^{#1}$}
\def\E#1 {E{#1}}
\def\P#1,{$\nH2\TK~=~#1\times~10^4\pccc$~K}

\def\H3{\mbox{H$_3$}}
\def\Uhcop{\mbox{$\Upsilon_\hcop$}} 
\def\NHIcl{\mbox{N(HI)$_{\rm cl}$}} 
\def\L21{\mbox{{$\lambda$21cm}}}
\def\Whcop{\mbox{$\Upsilon_{{\rm HCO}\p}$}}

\newcommand{\emm}[1]{\ensuremath{#1}}   % ensures math mode.
\newcommand{\emr}[1]{\emm{\mathrm{#1}}} % uses math roman fonts.
\newcommand{\hcop}{\emr{HCO^+}} 
\newcommand{\HI}{\emr{H\textsc{i}}}

\newcommand{\X}[1]{\emm{X_\emr{#1}}}
\newcommand{\XCO}{\X{CO}}
\newcommand{\W}[1]{\emm{{\rm W}_\emr{#1}}}
\newcommand{\WCO}{\W{CO}}

\begin{document}

\title{The Dark Neutral Medium is (Mostly) Molecular Hydrogen}

\author{ H. Liszt\inst{1} and M. Gerin\inst{2}}
\institute{
     National Radio Astronomy Observatory,
           520 Edgemont Road,
           Charlottesville, VA,
           USA 22903 
      \email{hliszt@nrao.edu}
\and
LERMA, Observatoire de Paris, PSL Research University CNRS Sorbonne Universit\'e \\
\email{maryvonne.gerin@obspm.fr}
%\and
%AIM, 
%CEA-IRFU/CNRS/Universit\'e Paris Diderot,
%D\'epartement d'Astrophysique, CEA,
%Saclay, 91191, Gif-sur-Yvette, France
%\email{isabelle.grenier@cea.fr}
}

 %
%\thesaurus{
%              09         % A&A Section 9
%              (09.01.1 %ISM: abundances
%               09.03.1 %ISM: clouds
%               09.13.2 %ISM: molecules
%               09.19.1 %ISM: structure
%               13.19.3 %Radio lines: interstellar)
%             }

\date{received \today}%
\offprints{H. S. Liszt}%
\mail{hliszt@nrao.edu}%
%
% \abstract{}{}{}{}{} 
% 5 {} token are mandatory
\abstract
  % context heading (optional) leave it empty if necessary  
{More gas is sometimes inferred in molecular cloud complexes  than is represented in HI 
or CO emission, and this is called dark neutral medium (DNM).}
% aims heading (mandatory)
{Our aim is to extend a study of DNM along 13 directions in the outskirts of 
 Chamaeleon by determining the atomic or molecular character of the DNM along 
 a larger sample of sightlines.}
% methods heading (mandatory)
{We acquired ALMA ground rotational state absorption profiles of \hcop\ and 
 other molecules toward 33 compact extragalactic continuum background sources 
 seen toward the Galactic anticenter, deriving N(\HH) = N(\hcop)/$3\times 10^{-9}$ as 
 before. We observed J=1-0 CO emission with the IRAM 30m telescope in directions 
 where \hcop\ was newly detected.} 
%results heading (mandatory)
 {\hcop\ absorption was detected in 28 of 33 new directions and CO emission 
 along 19 of those 28. 
 %Overall, \hcop\ has been detected in 41 of 46 directions
 %and absorption from \cch\ and HCN in 26/46 and 21/46 directions, respectively.
 The five sightlines lacking detectable \hcop\ have three times lower $<$\EBV$>$
 and  $<$N(DNM)$>$. Binned in \EBV, N(\HH) and N(DNM) are strongly 
 correlated and vary by factors of  50-100 over the observed range \EBV\ $\approx$ 
 0.05 - 1 mag, while N(HI) varies by factors of only 2-3. On average N(DNM) 
 and N(\HH) are well matched, and detecting \hcop\ absorption adds little to no 
 \HH\ in excess of the previously inferred DNM.  There are five cases 
 where  2N(\HH) $<$ N(DNM)/2 indicates saturation of the HI emission 
 profile. For sightlines with \WCO\ $\ge$ 1 K-\kms,\ the CO-\HH\ conversion factor 
 N(\HH)/\WCO\ $= 2-3\times10^{20}~\pcc$/(1 K-\kms) is higher than is derived
 from  studies of resolved clouds in $\gamma$-rays.}
%conclusions heading (optional), leave it empty if necessary
 {Our work sampled primarily atomic gas with a mean \HH\ fraction $\approx$ 1/3, 
 but the DNM is almost entirely molecular. CO fulfills its role as an \HH\ tracer 
 when its emission is strong, but large-scale CO surveys are not sensitive to \HH\ 
 columns associated with typical values N(DNM) = $2-6 \times 10^{20}\pcc$.
 Lower \XCO\ values  from $\gamma$-ray studies arise in
 part from different definitions and usage. 
%  \hcop\ absorption provides valuable information on molecular gas along
%  low-extinction lines of sight near the HI-\HH\ transition.
  Sightlines with \WCO\ $\ge$ 1 K-\kms\ represent 2/3 of the \HH\ detected in
 \hcop\ and detecting 90\% of the \HH\  would require detecting CO at 
 levels \WCO\  $\approx$ 0.2-0.3 K-\kms .}


\authorrunning{L\&G} \titlerunning{DNM, dark and molecular gas}

\maketitle

%\comment{%
%  Tutorial of natbib:
%  \begin{itemize}
%  \item \citet{PetLuc+08}
%  \item \citep{PetLuc+08}
%  \item \citep[before][]{PetLuc+08}
%  \item \citep[][after]{PetLuc+08}
%  \item \citep[before][and after]{PetLuc+08}
%  \end{itemize}
% I made the changes where needed, \ie{} double parenthesis.}
%

%s1
\section{Introduction}
\label{section:Section1}
Between the atomic and molecular interstellar gases there is a transition
 that is imperfectly traced in  $\lambda 21$cm HI and/or
$\lambda 2.6$mm CO emission. Gas that is not well represented
in one or both tracers is described as dark neutral medium (DNM)
\citep{GreCas+05,Pla15Cham,RemGre+17,RemGre+18}, and the atomic or molecular 
character of the DNM can only be decided by appealing to other information.

In the outskirts of Chamaeleon, we used  millimeter-wave \hcop\ absorption  to show that 
the DNM was almost exclusively molecular, even while the gas as a whole  
was mostly atomic \citep{LisGer+18}. \hcop\ was detected along all 13 sampled 
sightlines, of which 12 lacked detectable CO emission at a level of 1.5 K-\kms.  
The amount of inferred \HH\ was comparable to that of the DNM 
(confirming the amount of the detected DNM), and   only for three
sightlines did it seem likely that saturation of the HI emission profile 
could account for much of the DNM. Subsequent detections of CO absorption 
along six directions by \cite{LisGer+19} showed that the lack of CO emission was 
due to low CO column densities.  The linewidth of CO absorption was shown to 
be smaller than that of its parent molecule \hcop, illustrating the complex
chemical nature of the CO formation process in diffuse gas.

In this work we extend the analysis of the composition of DNM to 33 sightlines 
in directions toward the Galactic anticenter \citep{RemGre+17,RemGre+18} and consider the
larger sample of 46 sightlines.  The present study employs a relatively small number  
of diffuse (\AV\ $\le 1$ mag) and translucent (\AV\ $\le 3$ mag) sightlines where 
background sources are serendipitously available, and is not 
a revision of the earlier study, whose derived DNM column densities and other
results are assumed here.  Rather, we use the newly derived \HH\ abundances, 
independent of the presence of CO emission, to consider such topics as the 
atomic or molecular character of the DNM, the degree to which the \L21\ HI profile 
represents N(HI) and the influence of the CO-\HH\ conversion factor on the DNM 
determination.   

Section~\ref{section:Section2} summarizes the observational
material considered here and Section~\ref{section:Section3} compares the prior results for DNM
with atomic hydrogen measured in the  \L21\ emission and \HH\ as sampled
by its proxy 89 GHz absorption of \hcop.  The role of CO emission
data is explored in Section~\ref{section:Section4} where the CO-\HH\ conversion factor
is derived. Section~\ref{section:Section5} presents a summary, with conclusions and a discussion of 
later work.

%s2
\section{Observations and data reduction}
\label{section:Section2}
The methods of work were described in \cite{LisGer+18} and are not repeated
in full detail here.  From the existing DNM analyses of
\cite{Pla15Cham} and \cite{RemGre+17,RemGre+18}
we take N(DNM) and \L21\ HI column densities N(HI) and \NHIcl\ representing respectively 
the line profile integral integrated over all velocities and the column density that 
is associated with the cloud and/or kinematic features hosting the neutral gas and DNM,
as derived by a decomposition that accounts for the compound nature of HI emission profiles 
from individual clouds.

To the existing analyses we add ALMA observations of absorption from
\hcop\ and profiles of $\lambda$2.6mm CO emission from the IRAM 30m 
telescope, as described below.

%ss21
\subsection{Millimeter-wave absorption}
The new absorption data discussed here are spectra of \hcop\ in 33 directions along
Galactic anticenter sightlines toward ALMA phase calibrators as given in the Tables here
and projected on sky maps in Figure~\ref{fig:Figure1}. As before,
the \hcop\ spectra were acquired with a spectral resolution of 0.205 \kms\
sampled at 0.102 \kms\ intervals without the so-called spectral averaging
that bins data. We also acquired  spectra of HCN, \cch, HCO, and several 
isotopologs.  Counts of detections are shown in Figure~\ref{fig:Figure2}, but results for
species other than \hcop\  will  be discussed in detail in later work.
Statistics of the \hcop\ detections are shown in Figure~\ref{fig:FigureA.1}.
Spectra in the 28 directions with detectable \hcop\ absorption are shown in 
Figure~\ref{fig:FigureB.1}.

%ss22
\subsection{IRAM 30m CO emission}

We took frequency-switched CO emission profiles at the IRAM 30m telescope 
in August 2019 toward the 28 anticenter  sightlines with detected \hcop\ 
absorption. The CO spectra are shown in Figure~\ref{fig:FigureB.1}.
These CO data were taken as five-point maps toward the target and displaced
by 2 HPBW (2x22\arcsec) in the four cardinal directions, using a pointing
pattern that was designed to detect \hcop\ and HCN emission in the presence 
of spectral line absorption in the target direction.
For the \hcop\ emission that will be discussed in subsequent work,
the \hcop\ emission profile is formed by averaging spectra along the four 
outlying directions.  For CO the present results use the average of all five
pointings because contamination by absorption is not measurable given
the strength of the emission and/or the continuum.  The results
are presented on the main beam antenna temperature scale that is
native to the 30m telescope.

%ss23
\subsection{Conversion from integrated HCO\p\ absorption to N(H$_2$)}

The suitability of \hcop\ as a proxy for \HH\ in diffuse molecular gas 
was explored extensively in \cite{LisGer23} (hereafter   
Paper 1). As in our earlier work, we use N(\HH) = N(\hcop)/$3\times10^{-9}$
and N(\hcop) = $1.10\times 10^{12}$\Uhcop\ where \Uhcop\
is the integrated \hcop\ optical depth expressed in \kms.

%1
\begin{figure*}
\includegraphics[height=8.4cm]{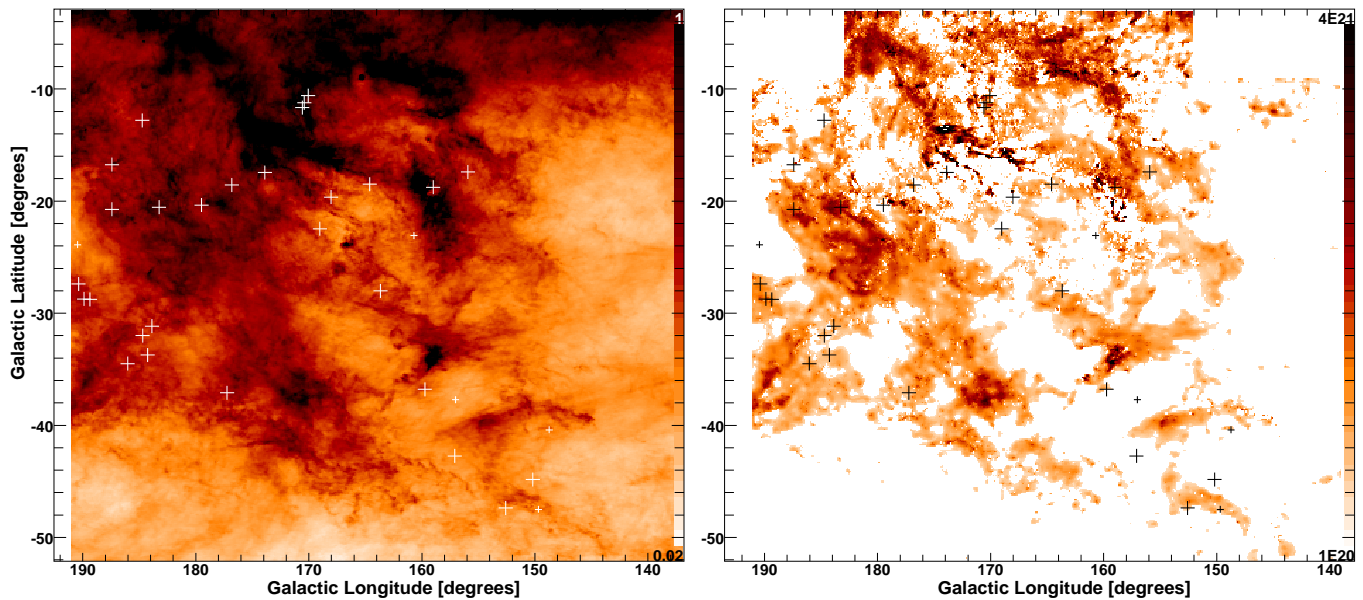}
  \caption[]{Sky maps of the observed sightlines. Left: Positions of the background sources observed here
are projected against a map of \EBV\ from the work of \cite{SchFin+98}. The 
sightlines lacking detected \hcop\ absorption (largely due to weak continuum) 
are indicated with smaller symbols. Right: Same as left,  but shown against a map 
of N(DNM) from \cite{RemGre+17,RemGre+18}.}
\label{fig:Figure1}
\end{figure*}

%ss24
\subsection{Reddening and dust optical depth}

The 6\arcmin\ resolution dust-emission maps scaled to optical
reddening \EBV\ by \cite{SchFin+98} are cited here. Those 
reddening values can be converted to Planck 353 GHz dust optical 
depth using the relationship established by \cite{PlaXI} between
the 353 GHz dust optical depth and the \EBV\ values of \cite{SchFin+98}, 
\EBV/$\tau_{353} = (1.49\pm0.03) \times 10^4$ mag.

%ss25
\subsection{Conventions}

Velocities presented with the spectra are taken with respect to the kinematic 
definition of the Local Standard of Rest.  N(H) is the column 
density of H nuclei detected in neutral atomic and molecular form, 
N(H) = N(HI)+2N(\HH), and the molecular hydrogen and DNM fractions
2N(\HH)/N(H) and N(DNM)/N(H) are respectively represented by \fH2\ and \fDNM.
The integrated absorption of the \hcop\ profile in units of \kms\ is denoted 
by \Whcop\ and similarly for other species.  The integrated emission of the 
J=1-0 CO line is denoted by \WCO\ with units of K-\kms\ 
and the  CO-\HH\ conversion factor N(\HH)/\WCO\ is denoted by \XCO.
Where reference is made to a typical Galactic or standard CO-\HH\ conversion 
factor, the value N(\HH)/\WCO\ $= 2\times10^{20}~\pcc$/(1 K-\kms) should
be understood, as summarized in Table E.1 of \cite{RemGre+17}. 

%ss26
\subsection{Overview}

Observational results of importance to this work are given in Table~\ref{table:Table1}
and Table~\ref{table:Table2}, and summaries of mean properties of various  subsamples are presented in 
Table~\ref{table:Table3}. Some statistics of the continuum targets and noise in the detections of
\hcop\ absorption are discussed in Appendix~\ref{section:AppendixA} and the CO emission and
\hcop\ absorption line profiles are shown in Figure~\ref{fig:FigureB.1} and discussed in 
Appendix~\ref{section:AppendixB}.  

%s3
\section{HI, DNM, and \HH\ sampled in HCO$^+$ absorption}
\label{section:Section3}
Figure~\ref{fig:Figure1} shows the new Galactic anticenter sightlines projected on maps of \EBV\ 
and N(DNM). The region at $b < -30$\degr\ is described by \cite{RemGre+17} as
Cetus, and that at $b > - 22$\degr\ as Taurus-Main. California lies to the 
north around $b = -10$\degr\ and Perseus is near (l,b) = 160\degr,-20\degr. 
Taurus-North overlays most of the map region.  The individual subregions
are not discussed   here owing to the sparse sampling.

Counts of sources and detected molecular tracers binned in 
0.05 mag  intervals of reddening are shown in Figure~\ref{fig:Figure2}.  \hcop\ is the molecular 
tracer detected most commonly (28 of 33 anticenter sightlines and 41 of 46 overall), 
followed by \cch\ (26 total) and HCN (21 total) .  The properties of molecular 
absorption line tracers other than \hcop\ will be discussed in later work.  

The Chamaeleon and Galactic anticenter subsamples have the same mode at 
\EBV\ = 0.25 - 0.3 mag in Figure~\ref{fig:Figure2}, but the anticenter subsample
has a high-reddening tail at \EBV\ $> 0.5$ mag that is absent in Chamaeleon.
The Galactic anticenter sample has higher mean extinction and molecular fraction. 
Overall (see Table~\ref{table:Table3}) the anticenter sources have the following characteristics: 
\begin{itemize}
\item{33\% higher $<$\EBV$>$ = 0.36 vs 0.27 mag};
\item{Same $<$N(HI)$>$ = $1.3\times 10^{21}\pcc$ and \fDNM\  =  0.12};
\item{80\% higher $<$\NHIcl$>$ = 1.12 vs 0.62 $\times 10^{21}\pcc$};
\item{2 times higher $<$2N(\HH)$>$ = 0.80 vs 0.39 $\times 10^{20}\pcc$};
\item{65\% higher \fH2\ = 0.38 vs 0.23};
%\item{Higher fraction of detected HCN, 17/33 vs 4/13}
\item{Higher fraction of detected CO, 9/33 (18/33 in the new work) vs 1/13};
\item{Higher incidence of nondetection of \hcop, 5/33 vs 0/13, in some
cases due to low flux,  see Figure~\ref{fig:FigureA.1}.}
\end{itemize}

The DNM fraction is nearly the same in the two samples (0.12-0.13) and much smaller 
(0.06) in both subsamples when CO emission was detected at the level of 1 K-\kms.
The DNM fraction is also noticeably smaller when molecular gas sampled in
\hcop\ absorption is absent, and similarly  at \EBV\ $<$ 0.2 mag.
The strongest variations in \fDNM\ in Table~\ref{table:Table3} arise in selections based on 
the strength of CO emission and are discussed in Section~\ref{section:Section4}.

%ss31
\subsection{Quantitative summary results}
\label{section:Section3.1}
\begin{figure}
\includegraphics[height=11cm]{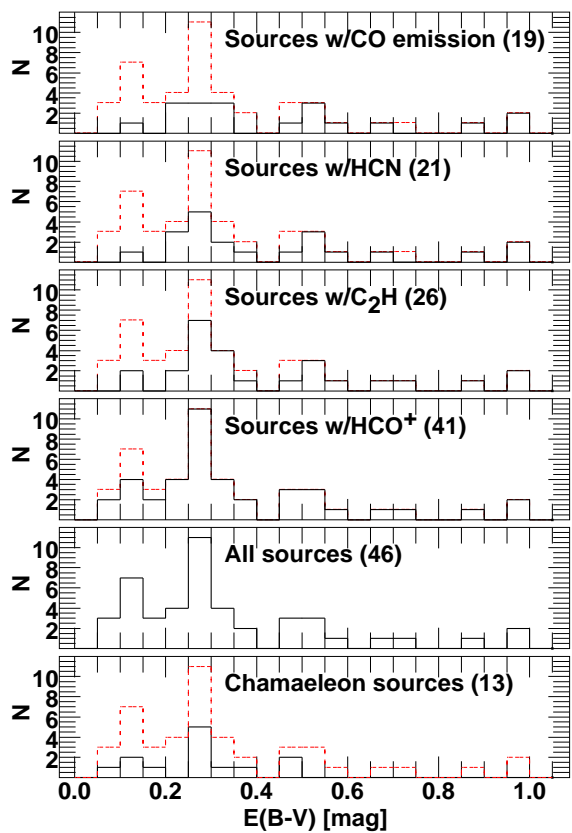}
  \caption[]{Counts of sightlines and detected tracers binned in 0.05 intervals 
 of reddening, \EBV. The histogram of all sources is overlaid (dashed  red)
in the other panels.}
\label{fig:Figure2}
\end{figure}

\begin{figure*}
\includegraphics[height=7.75cm]{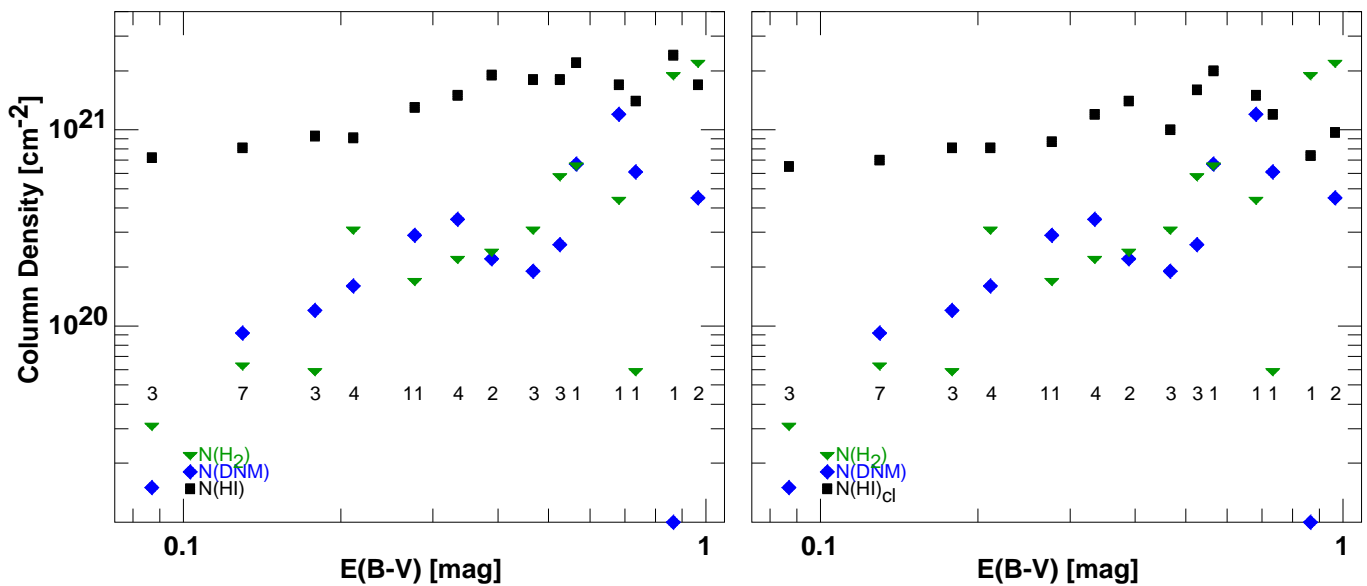}
\caption[]{
Trends in mean N(HI), N(\HH), and N(DNM). The results are binned in 0.05 mag intervals of 
\EBV. Left: N(HI) is calculated over the whole line profile. Right: N(HI) = \NHIcl\ the HI 
column density associated with the clouds studied here. The number of sightlines
in each bin is shown in both panels.}
\label{fig:Figure3}
\end{figure*}

Figure~\ref{fig:Figure3} shows an overview of  trends in means of N(HI), N(DNM), and N(\HH) 
using data binned in 0.05 mag intervals of reddening as in Figure~\ref{fig:Figure2}, plotted 
horizontally at the mean \EBV\ in every bin.  The counts in each bin are shown 
and the bins at \EBV\ $>$ 0.5 mag are occupied by only one or two sightlines. 
The association of DNM with molecular gas is unmistakable.

The mean HI column density varies by only a factor of 3 over a wider range \EBV\ = 0.09 - 1 
mag.  By contrast, means of N(NDM) and N(\HH) vary by factors of 50-100. 
The values of
$<$N(DNM)$>$ and $<$N(\HH)$>$ are comparable and increase steadily for \EBV\ $\la$ 0.7 mag, 
beyond which $<$N(DNM)$>$ either fails to increase or falls: the CO emission tracer used to
derive N(DNM) was used effectively to account for H nuclei in \HH. 
The cloud-associated mean \NHIcl\ declines at \EBV\ $\ge 0.7$ 
mag as \HH\ sequesters H nuclei. The value of $<$N(\HH)$>$ increases up to the bins at highest \EBV\ 
where $<$N(\HH)$>$ $\approx$ $<$N(HI$>$) $\approx 2-3$ $<$\NHIcl$>$ and whole sightlines 
are dominated by molecular gas.

%2

The correlation between DNM and \HH\ is remarkable considering the vastly 
different scales on which \EBV, N(HI), and N(NDM) are measured (6-20\arcmin), as 
contrasted with N(\HH) $\propto$ \Uhcop\ that is derived on submilliarcsecond 
scales from absorption against scatter-broadened compact millimeter-wave ALMA calibrator 
continuum  sources.  The same kind of correlation of tracers observed on different 
angular scales occurs in the  correlation of integrated $\lambda$21cm HI 
optical depth with \EBV\ \citep{Lis21}.

When \hcop\ is  not detected (5/46 directions) the following characteristics are found
compared to sightlines with detected \hcop:  
\begin{itemize}
\item{3 times lower $<$\EBV$>$ = 0.13 mag;} 
\item{1.3 times higher $<$N(H)$>$/$<$\EBV$>$ = $8\times10^{21} \pcc$ mag$^{-1}$};
\item{3.5 times lower $<$N(DNM)$>$ $= 8\times10^{19} \pcc$};
\item{2 times lower $<$\fDNM$>$ = 0.07};
\item{no DNM in 3/5 cases vs. 15/46 overall}.
\end{itemize}

Instead, using 3$\sigma$ upper limits for N(\HH) we find:
\begin{itemize}
\item{$>$5 times lower $<$2N(\HH)$> \le 1.5\times10^{20}\pcc$} and 
\item{$>$2.5 times lower $<$\fH2$> \le 0.14$}
\end{itemize}

The  ensemble of sightlines 
with N(DNM) $< 5\times 10^{19}\pcc$ does not differ by more than 10-20\% from other 
sightlines in any mean property not involving N(DNM).
The variations in N(DNM) and N(\HH) stand in great  contrast with N(HI) that 
increases only by a factor of  3 across the figure. The rise in N(\HH)
accounts for the failure of N(HI) to increase at high \EBV,\ but this could in 
principle also be explained by increasing saturation of the HI emission.  As 
noted in \cite{Pla15Cham} and further refined by \cite{RemGre+18}, the global
DNM analysis finds the best fit with the optically thin estimate of N(HI), 
implicitly leaving a bigger overall role for molecular gas than for saturation 
of the HI profile.

%ss32
\subsection{DNM and HI}
\label{section:Section3.2}
%3

Shown in Figure~\ref{fig:Figure4} is the variation of $<$N(DNM)$>$ with $<$N(HI)$>$ and 
$<$\NHIcl$>$. Chamaeleon differs from the Galactic anticenter 
in having more extraneous atomic gas along the line of sight and smaller values 
of cloud-associated gas. The cloud-associated HI in Chamaeleon is clustered 
at the low end of the range, even though the samples are evenly matched in the 
span of their total HI column density and have the same mean N(HI) in Table~\ref{table:Table3}.

Sightlines lacking DNM are present at all N(HI), but N(DNM) is uniformly small 
along sightlines at the highest column densities N(HI) $\ga 2.3\times10^{21}\pcc$ 
where CO emission is more commonly observed.
%There are no sightlines with $<$\NHIcl$>$ $> 2\times 10^{21}\pcc$, the column
%density that is familiarly associated with 1 magnitude of optical extinction.
The sources with N(DNM) $> 6\times 10^{20}\pcc$ and high N(DNM)/N(HI) ratios
are perhaps the most obvious candidates for hosting significant quantities of
DNM in atomic form as the result of saturation of the HI profile, but only two 
of these actually have small contributions from N(\HH). Assessing the DNM
contribution arising from possible saturation of the HI profile is discussed in 
Section~~\ref{section:Section3.3} where it is seen that the strongest cases of atomic DNM actually have 
more modest N(DNM).    

%ss33
\subsection{DNM, \EBV,\ and N(\HH)}
\label{section:Section3.3}
As with HI, sightlines lacking DNM are present over the full range of \EBV\ in 
Figure~\ref{fig:Figure5},  top, but they are predominant at smaller N(\HH).
Sightlines with 
2N(\HH) $\la 10^{20}\pcc$ lack DNM. Along sightlines with column densities too
low to host molecular gas, the atomic gas is well represented by the optically
thin estimates of N(HI) used in the DNM analysis.

Care must be taken in interpreting the relationship of DNM and N(\HH) in Figure~\ref{fig:Figure5} 
because the DNM in part represents material that is missing in the CO emission tracer 
while some of the \HH\ traced by \hcop\ is actually visible in CO emission.  To minimize 
this cross-contamination, Figure~\ref{fig:Figure6} shows a plot of N(DNM) vs. N(\HH) for 
sightlines lacking a detection of CO, using the more sensitive IRAM data for the Galactic 
anticenter subsample. Included are all but one (12/13) of the Chamaeleon sightlines and 
60\% of those toward the anticenter.

Sightlines lacking DNM in the absence of CO emission in Figure~\ref{fig:Figure6} are almost 
entirely confined at 2N(\HH) $\la 2\times 10^{20}\pcc$, so the detections of \hcop\ 
absorption do not imply much additional gas along sightlines where DNM was 
not found.  Missing from Figure~\ref{fig:Figure6} are sightlines with 
2N(\HH) $\ga 10^{21}\pcc$ corresponding to \WCO = 2.5 K-\kms\ for
the Galactic CO-\HH\ conversion factor.  

There are two or three cases in each subsample where N(DNM)/2N(\HH) $\ga 2$ and 
saturation of the \L21\ emission profile may be important.  Most of these
sightlines have modest values N(DNM) $\approx 4-5\times 10^{20}\pcc$ and
are not the sightlines with high N(DNM) that were singled out for discussion 
in Section~\ref{section:Section3.2}  when comparing N(DNM) and N(HI). 

There are also two or three sightlines 
at N(\HH) $> 6\times 10^{20}\pcc$ where N(DNM)/2N(\HH) $\la 2$. 
Overall $<$2N(\HH)$>$ = 1.3$<$N(DNM)$>$ for sources lacking CO emission, 
so \HH\ accounts for DNM without adding extra ``dark'' gas, and the DNM is mostly 
molecular.  As before in Chamaeleon, the DNM is
largely molecular, even if the medium overall is predominantly atomic.

%s4
\section{The role of CO emission}
\label{section:Section4}
The CO emission that was detected in previous studies used to determine
N(DNM) along our sightlines was well above 1 K-\kms\ in every case (Table~\ref{table:Table2}), 
with  $<$\WCO$>$ = 4.9K-\kms\ (Table~\ref{table:Table3}). This is very nearly the same mean as 
for the  sample of sightlines with \WCO\ $\ge 1$K-\kms\ observed at the IRAM 
30m, 5.5 K-\kms\ (see Table~\ref{table:Table2}). Compared to the overall average, the sightlines with 
$<$\WCO$>$ $\ge$ 1K-\kms\ have the following characteristics: 
\begin{itemize}
\item{50-100\% higher $<$\EBV$>$ = $0.5-0.7$ mag};
\item{3-4 times higher $<$2N(\HH)$>$ = $2-3\times 10^{21}\pcc$}; 
\item{2 times higher $<{\rm f}_{\HH}>$ = 0.6};
\item{2 times lower $<$\fDNM$>$ = 0.06}. 
\end{itemize}

%4
\begin{figure}
\includegraphics[height=7.9cm]{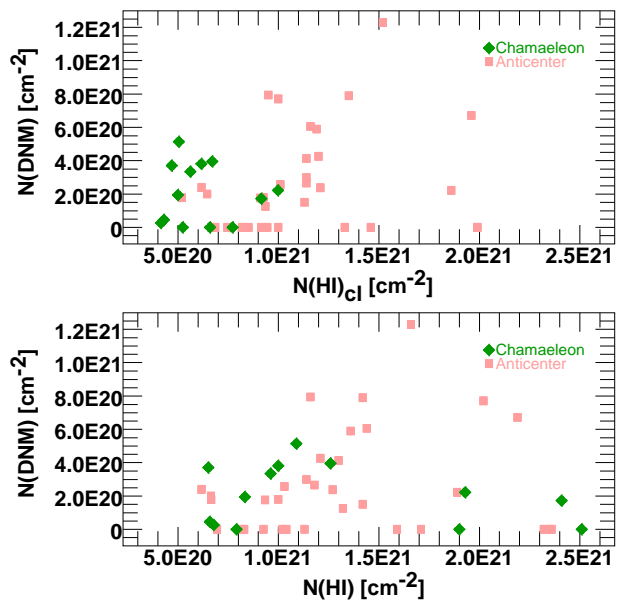}
  \caption[]{N(DNM) plotted against total sightline N(HI) (bottom) and 
   cloud-associated atomic hydrogen N(\HI)$_{cl}$ (top).}
\label{fig:Figure4}
\end{figure}

%5
\begin{figure}
\includegraphics[height=15cm]{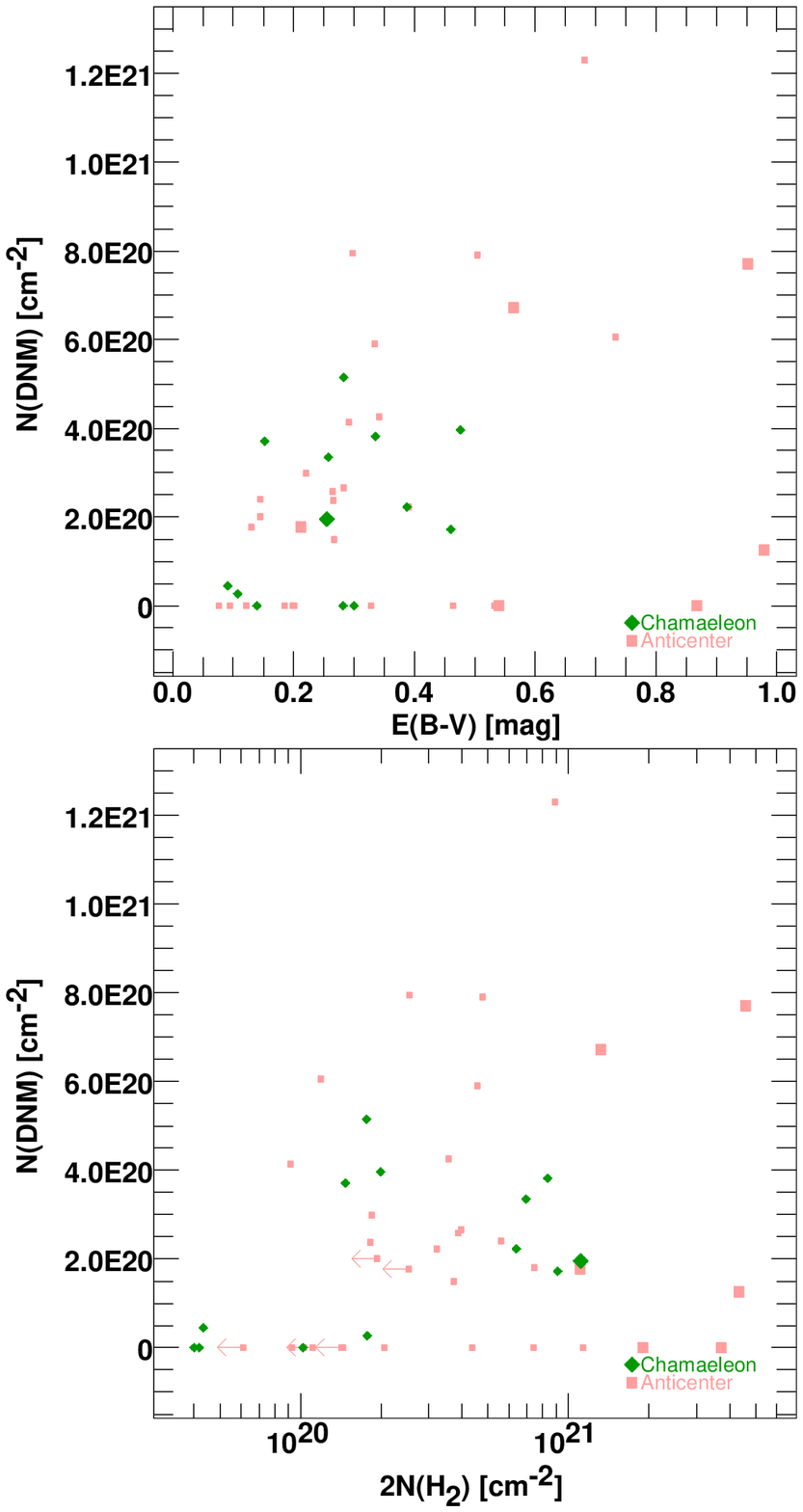}
  \caption[]{N(DNM) plotted against \EBV\ (top) and 2N(\HH) (bottom).
  Larger symbols represent sightlines where \WCO $\ga$ 1 K-\kms.}
\label{fig:Figure5}
\end{figure}

The 50\% smaller DNM fractions $<$\fDNM$>$ = 0.06 along samples of 
sightlines with $<$\WCO$>$ $\ge$ 1K-\kms\ indicate that CO emission 
is doing a good job of tracing \HH\ 
in gas where the molecular fraction is high \citep{Lis17CO}, 
CO emission is strong, and the cloud-associated 
HI fraction declines (Figure~\ref{fig:Figure2}). However, CO emission in general represents 
a small portion of the total molecular gas present along the sightlines in 
this work. For instance, $<$2N(\HH)$>$ = $7\times 10^{20}\pcc$ along 46 
sightlines, while $<$\WCO$>$ $\approx 5$ K-\kms\ along the 6-10 sightlines 
where $<$\WCO$>$ $\ge$ 1K-\kms\ (Table~\ref{table:Table3}). For a typical Galactic
CO-\HH\ conversion factor, the summed molecular gas column represented in
\hcop\ is three to four times that inferred from the summed CO emission. Most of the 
molecular gas in our sample was hidden in the DNM prior to our work and 
is still only revealed by the \hcop\ absorption, even with more sensitve CO 
observations. The IRAM 30m detections with $<$\WCO$>$ below 1K-\kms\ comprise 
less than 20\% of the total amount of CO emission.

%ss41
\subsection{The CO-H$_2$ conversion factor}
\label{section:Section4.1}
Comparison of the scaled \hcop\ absorption and IRAM 30 emission measurements
provides the most direct determination of the actual CO-\HH\ conversion factor
along the lines of sight studied in this work. Figure~\ref{fig:Figure7} shows the trends in
$<$N(\HH)$>$, $<$\WCO$>$, and $<$N(\HH)$>$/$<$\WCO$>$ binned in 0.05 mag increments
of \EBV\ as in Figure~\ref{fig:Figure3}. Included are the 28 Galactic anticenter sightlines where \hcop\
was detected, with \WCO\ taken at the 3$\sigma$ upper limit for sightlines where
CO emission was not detected with greater significance. Also included is
the sightline toward J1733 in Chamaeleon where CO emission was detected.
The $3\sigma$ upper limits \WCO\ $\le$ 1.5 K-\kms\ in Chamaeleon are not useful.    

Figure~\ref{fig:Figure7} illustrates the behavior of the CO-\HH\ conversion factor that is 
tabulated for different samples in Table~\ref{table:Table3}. The values of  $<$N(\HH)$>$ and $<$\WCO$>$ 
both increase steadily with \EBV\ in Figure~\ref{fig:Figure7}, but at different rates so that 
their ratio declines by a factor $\approx 7$ to 
$<$N(\HH)$>$/$<$\WCO$>$ $= 2.5-3\times 10^{20}$\HH$\pcc$(K-\kms)$^{-1}$ 
for \EBV\ $\ga$ 0.5 mag. Variations of similar magnitude in individual
diffuse and/or translucent MBM clouds were reported by \cite{MagOne+98}
and \cite{CotMag13}.

Mean values of N(\HH)/\WCO\ are $2-2.5\times 10^{20}$\HH$\pcc$(K-\kms)$^{-1}$ for 
the old and new samples with $<$\WCO$>$ $\ge 1$ K-\kms, increasing by factors of  2-3 
for the samples with IRAM 30m detections below 1K-\kms\ and upper limits. Also see the 
inset in Figure~\ref{fig:Figure7} on this point.. 

%6
\begin{figure}
\includegraphics[height=8cm]{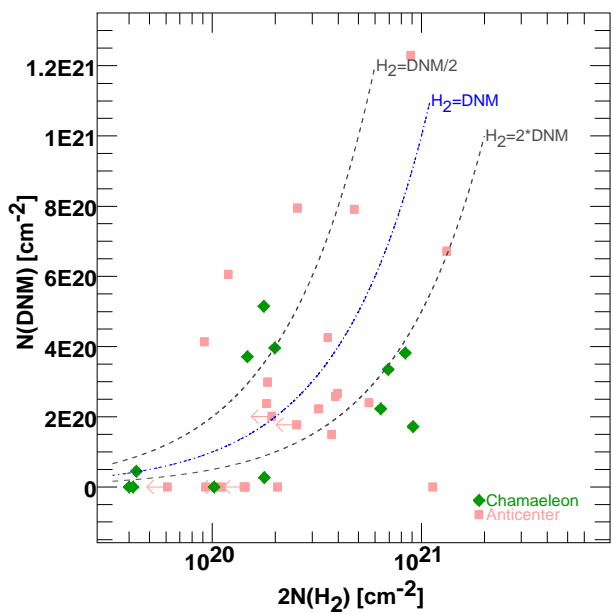}
  \caption[]{N(DNM) plotted against N(\HH) for sightlines lacking
detected CO emission in the analysis of \cite{RemGre+18}. Shown
are loci at which the number of H nuclei  in \HH\ is 50, 100, and
200\% of that in DNM.}
\label{fig:Figure6}
\end{figure}

%7
\begin{figure}
\includegraphics[height=7.6cm]{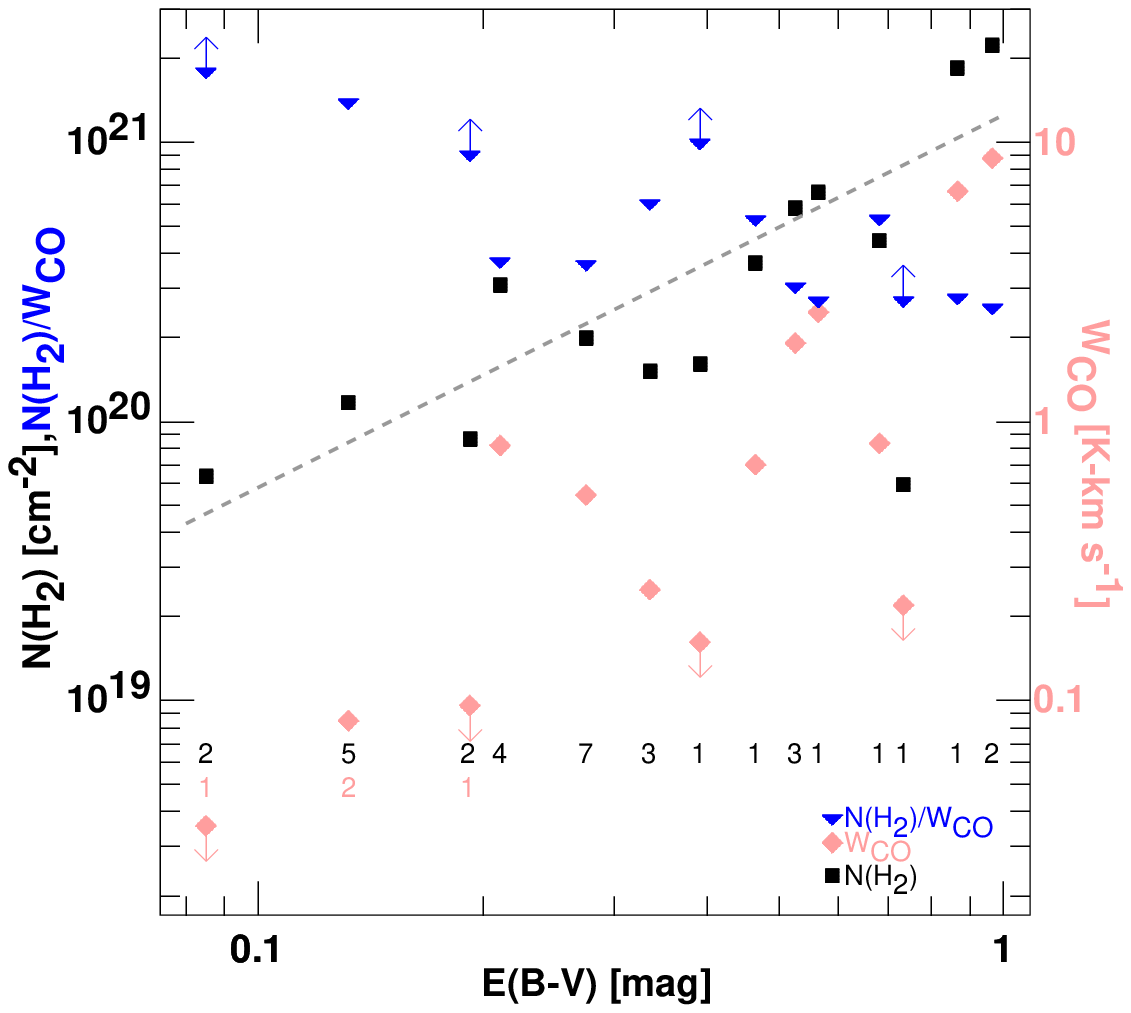}
\caption[]{
Trends in mean N(\HH) (black squares), \WCO\ (pink diamonds), and 
N(\HH)/\WCO\ (blue triangles). The  results are binned in 0.05 mag intervals of \EBV,\ as in Figure~\ref{fig:Figure3}. Bins in which all 
sightlines have only upper limits on \WCO\ are indicated by upper and lower limits.
The scales for N(\HH) $\pcc$ and N(\HH)/\WCO\ ($\pcc$ (K-\kms)$^-1$) are read on 
the left vertical axis and that for \WCO\ (K-\kms) on the  right. The number of sightlines 
in each bin is shown 
as in Figure~\ref{fig:Figure3} and the count is carried separately for \WCO\ at low \EBV. The light 
gray dashed line is a power-law fit N(\HH) $= 10^{21.0973}$\EBV$^{1.335}$.}
\label{fig:Figure7}
\end{figure}

The values of the CO-\HH\ conversion factor derived are comparable to those 
derived in extant Galactic-scale $\gamma$-ray analyses (Table E.1 in \citealt{RemGre+17}) 
when \WCO\ $\ga$ 1 K-\kms, but are uniformly larger
than those derived in cloud-level studies like the DNM analysis whose results are
summarized in Table 2 of \cite{RemGre+17}.  Those results ranged from 
$1-1.6\times 10^{20}$\HH$\pcc$(K-\kms)$^{-1}$ for determinations based on dust 
and from $0.44 - 1.00\times10^{20}$\HH$\pcc$ (K-\kms)$^{-1}$
for determinations based on gamma rays.

There is no contradiction with the larger values found in this work whose method
of studying widely separated, semi-randomly placed sightlines is similar to the larger 
scale studies. The strongly CO-emitting (\WCO\ $>$ 10 K-\kms) regions encountered 
in the cloud-level studies that have generally small values of \XCO\ (see Figure 13 
in \citealt{RemGre+18}) were not sampled here. 

There may also be a difference arising from the operational definition of the 
conversion factor.  The conversion factors in Table 2 in \cite{RemGre+17} are 
those that optimize the fit of the $\gamma$-ray or dust emissivity to a total 
hydrogen column density that is represented schematically as 
N(H) = N(HI)+N(DNM)+2N(\HH) = N(HI)+N(DNM)+2\XCO\WCO. After the analysis, there 
remains a DNM constituent whose molecular fraction is undetermined and not 
explicitly considered in the definition of the multiplier \XCO;  we
showed that the DNM is largely molecular.   By contrast, the
present analysis determines N(\HH) independent of CO emission and defines
\XCO\ = N(\HH)/\WCO. This N(\HH) includes the molecular component of the 
DNM that we took pains to consider separately in the discussion of Figure~\ref{fig:Figure6}.

%ss42
\subsection{Achievable limits and detection thresholds for CO, DNM, and \HH}
\label{section:Section4.2}
Sightlines in our study often had rms CO emission noise 
$\Delta$\WCO\ $\ga 1/3-1/2$ K-\kms\ in the prior DNM analysis (Table~\ref{table:Table2}). For a 
typical Galactic conversion 2N(\HH)/\WCO\ $= 4\times 10^{20}\pcc$ (K-kms)$^{-1}$, 
the equivalent $3\sigma$ threshold detection limits on the hydrogen column are 
2N(\HH) $= 4-6\times 10^{20}\pcc$, above the actual values of N(DNM) along 
most of the sightlines we   observed. By contrast, the median $3\sigma$ 
detection threshold from \hcop\ in our work is 2N(\HH) $> 1.1\times10^{20}\pcc$
and typical N(DNM) values are N(DNM) $\ga 2 \times 10^{20}\pcc$ (Figure~\ref{fig:Figure6}).

The effectiveness  of reducing the detection 
threshold of CO emission below 1 K-\kms\ can be assessed by examining 
the distribution of CO detections and upper limits in the IRAM 30m CO observations.  
The subsamples of IRAM CO observations summarized in Table~\ref{table:Table3} show 6 detections with 
\WCO\ $>$ 1 K-\kms, 12 detections with \WCO\ $<$ 1 K-\kms,\ and ten upper limits
at levels \WCO\ $< 0.1 - 0.2$ K-\kms\ (see Table~\ref{table:Table1} for values of upper limits
and Figure~\ref{fig:Figure3} for a graphical representation of the data). 
\hcop\ was detected along all of these sightlines, and the three subsamples
respectively represent fractions 0.66, 0.26, and 0.08 of the total amount of \HH. 

Thus, a survey with a detection limit of 1 K-\kms\ would detect approximately two-thirds 
of the molecular gas along these diffuse and/or translucent lines of sight, and a much 
increased effort to reduce the detection limit to 0.2 K-\kms\ might find another 
one-fourth of the \HH. Comparable reductions in the fraction of undetected
\HH\ with increasing CO sensitivity down to rms levels 
$\Delta$\WCO\ $\approx 0.1$ K-\kms\ were achieved by \cite{DonMag17}.
Missing one-third of the \HH\ at the 1 K-\kms\ threshold
is consistent with the dark gas fraction derived by \cite{WolHol+10} and 
\cite{GonMun+18}.

%ss43
\subsection{The wider view}
\label{section:Section4.3}
In Section~\ref{section:Section4.2} we note that two-thirds of the \HH\ was found along
the sightlines with \WCO\ $>$ 1 K-\kms\ and that a deeper 
survey with a detection limit of 0.2 K-\kms\ would have found another 25\% 
of the \HH. At this point we  can ask   how the results of this sparse sampling are reflected 
in the region as a whole. We drew a hull around the observed anticenter sightlines, 
as shown in Figure~\ref{fig:FigureC.1} 
and derived the pixel-by-pixel statistics for the contained 
area, the amount of material (taken proportional to \EBV), and the amount of \HH. For
\HH\  we used the N(\HH)-\EBV\ relationship N(\HH) $= 10^{21.0973}$\EBV$^{1.335}$ 
\footnote{\fH2\ = 2N(\HH)/N(H) = 0.4-0.5 at \EBV\ = 2 mag if 
N(H)/\EBV\ $= 6-8\times10^{21} {\rm mag}^{-1}$}  
and the fact that \WCO\ $\ga 1$ K-\kms\ at \EBV\ $\ga$ 0.4 mag or 
N(\HH) $\ga 4\times 10^{20}\pcc$ in Figure~\ref{fig:Figure7} (see also Figure 5 of Paper 1).

\begin{figure*}
\includegraphics[height=8cm]{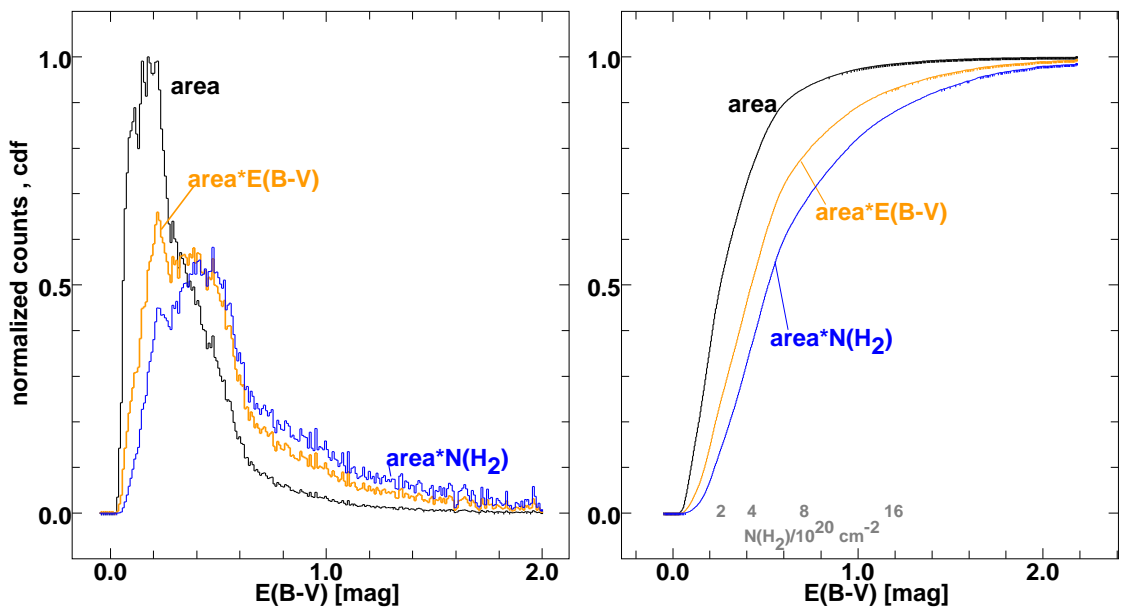}
\caption[]{
Probability distributions of  area, amount of material, and molecular hydrogen
with respect to \EBV. Left: Counts (probability density) scaled to the same area. 
Right: Cumulative  distribution function. N(\HH) is calculated  
using the mean power-law fit N(\HH) $= 10^{21.0973}$\EBV$^{1.335}$ derived
as shown in Figure~\ref{fig:Figure7}.}
\label{fig:Figure8}
\end{figure*}

Figure~\ref{fig:Figure8} shows the probability densities (at left) and cumulative distributions 
for the derived quantities. Reading values off the cumulative probability
distributions at right in Figure~\ref{fig:Figure8}, conditions with \EBV\ $\geq$ 0.4 mag, 
N(\HH) $\ga 4\times 10^{20}\pcc$ occur over one-fourth of the total area 
containing one-half of the total projected gas column and two-thirds of the 
\HH. Apparently, the sampled sightlines were representative of the region 
as a whole regarding the \HH\ distribution.

Sampling 90\% of the \HH\ would require detecting CO emission at \EBV\ $\ga 0.2$ 
mag where N(\HH) $\approx 2\times 10^{20}\pcc$. The sampling here is too sparse 
to derive an equivalent value of \WCO\ in Figure~\ref{fig:Figure7}, but the broader sample in 
Paper 1 suggests \WCO\ $\ga$ 0.3 K-\kms\ would be appropriate. Sightlines with 
\EBV\ $< 0.32$ mag or \AV\ $< 1$ mag comprise two-thirds of the area, one-third 
of the mass, and one-fourth of the \HH.

%8

%ss44
\subsection{Comparison with predictions and models}
\label{section:Section4.4}
For models, the critical factors in predicting the amount of dark gas are
the threshold where \WCO\ reaches the 1 K-\kms\ brightness level and the 
amount of material above the threshold.  In the regime of diffuse molecular gas
the integrated brightness of the J=1-0 line is determined almost exclusively by the 
CO column density, with \WCO\ (K-\kms) = N(CO)$/10^{15}\pcc$ for hydrogen 
number densities n(H) $= 64-500 \pccc$ \citep{Lis07CO,Lis17CO,HuSte+21} and
kinetic temperatures appropriate to the local thermal pressure.  This simple
proportionality between column density and brightness persists well beyond 
the CO column density where the J=1-0 transition becomes optically thick, owing 
to the strongly subthermal excitation \citep{GolKwa74}.

Thus, the CO chemistry dominates the CO brightness, and the sky map of CO emission
is a map of the chemistry. The observations can be reproduced by an ad hoc CO 
formation chemistry in which a fixed relative abundance of \hcop\ N(\hcop)/N(\HH) 
$= 3\times 10^{-9}$ thermally recombines to CO if the \HH\ formation and shielding 
of \HH\ and CO are treated self-consistently along with the heating and cooling
\cite{Lis17CO}.
However, networks of chemical reactions in quiescent gas may fail to form the 
observed quantity of \hcop\ and produce too little CO. This causes the CO
brightness to reach 1 K-\kms\ at overly large values of \EBV\ and N(\HH)
where shielding of CO by dust and \HH\ make up for the deficit in the CO
formation rate (e.g., \citealt{GonMun+18} Figure 7; \citealt{HuSte+21} Figure 9).

Models with a weak CO formation chemistry have an innate tendency to overestimate 
the amount of CO-dark gas, but the actual amount of CO-dark gas depends on the
distribution of material.  In practice the DNM fraction seen by \cite{RemGre+18}
varied from 0 to 0.3 (their Figure 8) and the sightlines sampled here had 
$<$\fDNM$>$ =  0.13.

%6,7,8 was here

%t1
\begin{table*}
\caption{Line-of-sight properties and new results for \hcop\ and CO }
\label{table:Table1}
{
\small
\begin{tabular}{lcccccccccc}
\hline
Source & RA(J2000)&Dec(J2000)& $l$ & $b$ &\EBV$^1$& 89 GHz flux&\Whcop $^2$&$\sigma$\Whcop $^3$&\WCO $^4$&$\sigma$\WCO $^5$ \\  
      &  hh.mmsss &dd.mmsss&degrees & degrees &mag &Jy & \kms & \kms & K-\kms& K-\kms \\  
\hline
J0203+1134& 2.03464& 11.34492&149.6826&-47.4992&0.144&0.126&$\leq$0.263&0.088& & \\
J0209+1352& 2.09357& 13.52045&150.1800&-44.8290&0.094&0.223& 0.20&0.050&$\leq$0.071&0.024\\
J0211+1051& 2.11132& 10.51348&152.5781&-47.3674&0.145&0.462& 0.76&0.029& 0.36&0.025\\
J0213+1820& 2.13105& 18.20255&148.7289&-40.4014&0.130&0.093&$\leq$0.345&0.115& & \\
J0231+1322& 2.31459& 13.22547&157.0917&-42.7380&0.121&0.430& 0.14&0.025&$\leq$0.064&0.021\\
J0242+1742& 2.42243& 17.42589&157.0180&-37.7033&0.077&0.227&$\leq$0.151&0.050& & \\
J0252+1718& 2.52077& 17.18427&159.7420&-36.7885&0.220&0.172& 0.25&0.077&$\leq$0.096&0.032\\
J0325+2224& 3.25368& 22.24004&163.6700&-28.0213&0.213&1.162& 1.01&0.017& 0.94&0.051\\
J0329+3510& 3.29154& 35.10060&155.9152&-17.4042&0.267&0.570& 0.51&0.032&$\leq$0.143&0.048\\
J0329+2756& 3.29577& 27.56155&160.7030&-23.0743&0.198&0.158&$\leq$0.193&0.064& & \\
J0334+0800& 3.34533&  8.00144&177.2396&-37.0871&0.391&0.150& 0.44&0.088&$\leq$0.162&0.054\\
J0336+3218& 3.36301& 32.18293&158.9998&-18.7650&0.733&1.689& 0.16&0.009&$\leq$0.219&0.073\\
J0356+2903& 3.56085& 29.03423&164.6120&-18.4927&0.212&0.139& 1.50&0.090& 1.62&0.042\\
J0357+2319& 3.57216& 23.19538&169.0302&-22.4661&0.185&0.224& 0.28&0.027&$\leq$0.192&0.064\\
J0400+0550& 4.00117&  5.50431&184.2710&-33.7266&0.266&0.159& 0.25&0.063&$\leq$0.172&0.057\\
J0401+0413& 4.01199&  4.13344&186.0261&-34.4947&0.341&0.405& 0.49&0.021& 0.19&0.048\\
J0403+2600& 4.03056& 26.00015&168.0260&-19.6483&0.201&0.331& 0.60&0.029& 0.62&0.067\\
J0406+0637& 4.06343&  6.37150&184.7075&-32.0009&0.283&0.264& 0.54&0.051& 0.63&0.040\\
J0407+0742& 4.07291&  7.42075&183.8723&-31.1558&0.265&0.387& 0.53&0.031& 0.19&0.042\\
J0426+0518& 4.26366&  5.18199&189.3631&-28.7705&0.291&0.516& 0.12&0.020&$\leq$0.170&0.057\\
J0426+2327& 4.26557& 23.27396&173.8881&-17.4457&0.539&0.304& 2.57&0.057& 4.84&0.045\\
J0427+0457& 4.27476&  4.57083&189.8857&-28.7306&0.335&0.414& 0.62&0.024& 0.39&0.045\\
J0437+2037& 4.31038& 20.37343&176.8096&-18.5565&0.532&0.217& 1.54&0.073& 0.67&0.026\\
J0431+1731& 4.31574& 17.31358&179.4942&-20.3579&0.464&0.104& 1.01&0.110& 0.70&0.049\\
J0433+0521& 4.33111&  5.21156&190.3730&-27.3967&0.298&4.911& 0.35&0.003&$\leq$0.109&0.036\\
J0437+2940& 4.37044& 29.40138&170.5818&-11.6609&0.979&0.059& 5.92&1.138&10.42&0.027\\
J0438+3004& 4.38049& 30.04455&170.4116&-11.2283&0.952&0.689& 6.25&0.038& 7.11&0.026\\
J0439+3045& 4.39178& 30.45076&170.0655&-10.5913&0.867&0.195& 5.05&0.082& 6.69&0.027\\
J0440+1437& 4.40211& 14.37570&183.2538&-20.5438&0.681&0.337& 1.21&0.031& 0.83&0.029\\
J0445+0715& 4.45014&  7.15539&190.4535&-23.8898&0.121&0.275&$\leq$0.083&0.028& & \\
J0449+1121& 4.49077& 11.21286&187.4274&-20.7365&0.504&0.521& 0.65&0.022& 0.23&0.033\\
J0502+1338& 5.02332& 13.38110&187.4143&-16.7456&0.564&0.271& 1.81&0.059& 2.46&0.031\\
J0510+1800& 5.10024& 18.00416&184.7304&-12.7895&0.328&2.411& 0.13&0.005& 0.17&0.036\\
\hline
\end{tabular}
\\
$^1$From \cite{SchFin+98} \\
$^2$Integrated optical depth or $3\sigma$ upper limit over a 3 \kms\ interval \\
$^3$Integrated emission from our results or $3\sigma$ upper limit over a 3 \kms\ interval \\
$^4$ rms of detected emission or rms over a 3\kms\ interval \\
}
\end{table*}

%Figure 8 was here 
%t2
\begin{table*}
\caption{Line-of-sight properties and derived quantities }
\label{table:Table2}
{
\small
\begin{tabular}{lccccccccc}
\hline
Source & E(B-V) & N(HI)$^1$ & \NHIcl $^1$ & 2N(\HH) $^2$& \WCO$^3$ & N(DNM) $^1$ & 2N(H$_2$)/N(HI) & N(DNM)/N(HI) & \XCO $^4$\\
& mag&$10^{21}\pcc$ & $10^{21}\pcc$ &  $10^{21}\pcc$&K-\kms & $10^{21}\pcc$  & \\
\hline
J0203+1134 & 0.144 & 0.664 & 0.646 & $\leq$0.193 &  & 0.201 & $\leq$0.290 & 0.303&  \\
J0209+1352 & 0.094 & 0.696 & 0.686 & 0.144 &  &  0.00 & 0.206 &  0.00&  \\
J0211+1051 & 0.145 & 0.618 & 0.617 & 0.561 &  & 0.240 & 0.908 & 0.388&  \\
J0213+1820 & 0.130 & 0.933 & 0.911 & $\leq$0.253 &  & 0.177 & $\leq$0.271 & 0.190&  \\
J0231+1322 & 0.121 & 0.926 & 0.852 & 0.103 &  &  0.00 & 0.111 &  0.00&  \\
J0242+1742 & 0.077 & 0.820 & 0.819 & $\leq$0.111 &  &  0.00 & $\leq$0.135 &  0.00&  \\
J0252+1718 & 0.220 & 1.142 & 1.139 & 0.184 &  & 0.299 & 0.161 & 0.262&  \\
J0325+2224 & 0.213 & 0.999 & 0.926 & 0.742 &  2.82 & 0.180 & 0.743 & 0.180& 1.314 \\
J0329+3510 & 0.267 & 1.418 & 1.131 & 0.372 &  & 0.150 & 0.263 & 0.106&  \\
J0329+2756 & 0.198 & 1.129 & 1.002 & $\leq$0.142 & $\leq 1.97$ &  0.00 & $\leq$0.126 &  0.00&  \\
J0334+0800 & 0.391 & 1.893 & 1.855 & 0.322 &  & 0.223 & 0.170 & 0.118&  \\
J0336+3218 & 0.733 & 1.435 & 1.165 & 0.119 & $\leq$2.60 & 0.606 & 0.082 & 0.423&  \\
J0356+2903 & 0.212 & 0.666 & 0.520 & 1.099 &  1.40 & 0.178 & 1.650 & 0.267& 3.931 \\
J0357+2319 & 0.185 & 1.022 & 0.944 & 0.205 &  $\leq$0.79 &  0.00 & 0.200 &  0.00&  \\
J0400+0550 & 0.266 & 1.266 & 1.210 & 0.182 &  & 0.238 & 0.144 & 0.188&  \\
J0401+0413 & 0.341 & 1.209 & 1.202 & 0.357 &  & 0.426 & 0.296 & 0.352&  \\
J0403+2600 & 0.201 & 0.830 & 0.667 & 0.438 &  1.99 &  0.00 & 0.528 &  0.00& 1.101 \\
J0406+0637 & 0.283 & 1.182 & 1.143 & 0.397 &  & 0.266 & 0.335 & 0.225&  \\
J0407+0742 & 0.265 & 1.033 & 1.006 & 0.387 &  & 0.258 & 0.374 & 0.250&  \\
J0426+0518 & 0.291 & 1.305 & 1.141 & 0.091 &  $\leq$1.40 & 0.414 & 0.070 & 0.317&  \\
J0426+2327 & 0.539 & 1.586 & 1.331 & 1.887 &  5.78 &  0.00 & 1.190 &  0.00& 1.632 \\
J0427+0457 & 0.335 & 1.361 & 1.192 & 0.457 &  2.49 & 0.590 & 0.336 & 0.433& 0.918 \\
J0437+2037 & 0.532 & 2.324 & 1.992 & 1.130 &  $\leq$1.61 &  0.00 & 0.486 &  0.00&  \\
J0431+1731 & 0.464 & 1.714 & 1.457 & 0.739 &  3.09 &  0.00 & 0.431 &  0.00& 1.198 \\
J0433+0521 & 0.298 & 1.159 & 0.950 & 0.255 &  $\leq$1.48 & 0.795 & 0.220 & 0.686&  \\
J0437+2940 & 0.979 & 1.320 & 0.936 & 4.345 & 11.98 & 0.126 & 3.292 & 0.095& 1.814 \\
J0438+3004 & 0.952 & 2.019 & 1.002 & 4.586 &  7.70 & 0.771 & 2.271 & 0.382& 2.979 \\
J0439+3045 & 0.867 & 2.361 & 0.745 & 3.701 &  9.42 &  0.00 & 1.567 &  0.00& 1.964 \\
J0440+1437 & 0.681 & 1.659 & 1.524 & 0.888 &  $\leq$1.97 & 1.230 & 0.535 & 0.741&  \\
J0445+0715 & 0.121 & 1.044 & 0.915 & $\leq$0.061 &  &  0.00 & $\leq$0.058 &  0.00&  \\
J0449+1121 & 0.504 & 1.416 & 1.347 & 0.476 &  & 0.791 & 0.336 & 0.558&  \\
J0502+1338 & 0.564 & 2.188 & 1.957 & 1.324 &  $\leq$1.90 & 0.672 & 0.605 & 0.307&  \\
J0510+1800 & 0.328 & 2.347 & 1.986 & 0.092 &  $\leq$2.00 &  0.00 & 0.039 &  0.00&  \\
\hline
J0942-7731 & 0.336 & 1.000 & 0.617 & 0.837 &  $\leq$1.50 & 0.382 & 0.838 & 0.382&  \\
J1058-8003 & 0.152 & 0.651 & 0.470 & 0.147 & `` & 0.371 & 0.226 & 0.570&  \\
J1136-6827 & 0.460 & 2.407 & 0.915 & 0.910 & `` & 0.172 & 0.378 & 0.071&  \\
J1145-6954 & 0.387 & 1.925 & 0.998 & 0.638 & `` & 0.223 & 0.331 & 0.116&  \\
J1147-7935 & 0.300 & 2.506 & 0.773 & 0.040 & `` &  0.00 & 0.016 &  0.00&  \\
J1152-8344 & 0.283 & 1.088 & 0.505 & 0.176 & `` & 0.515 & 0.162 & 0.473&  \\
J1224-8313 & 0.257 & 0.962 & 0.563 & 0.693 & `` & 0.335 & 0.720 & 0.348&  \\
J1254-7138 & 0.282 & 1.902 & 0.660 & 0.102 & `` &  0.00 & 0.053 &  0.00&  \\
J1312-7724 & 0.476 & 1.257 & 0.671 & 0.199 & `` & 0.396 & 0.158 & 0.315&  \\
J1550-8258 & 0.107 & 0.679 & 0.416 & 0.177 & `` & 0.027 & 0.261 & 0.039&  \\
J1617-7717 & 0.091 & 0.659 & 0.431 & 0.043 & `` & 0.045 & 0.065 & 0.068&  \\
J1723-7713 & 0.255 & 0.834 & 0.500 & 1.105 &  2.40 & 0.195 & 1.325 & 0.234& 2.302 \\
J1733-7935 & 0.139 & 0.792 & 0.525 & 0.041 &  $\leq$ 1.50 &  0.00 & 0.052 &  0.00&  \\
\hline
\end{tabular}
\\
$^1$ HI and DNM column densities are from \cite{Pla15Cham} and \cite{RemGre+17,RemGre+18} \\
$^2$ N(\HH) = N(\hcop)/$3\times10^{-9}$ \\
$^3$ Integrated J=1-0 CO emission used by \cite{Pla15Cham} and \cite{RemGre+17,RemGre+18} \\
$^4$ \XCO\ = N(\HH)/\WCO\
}
\end{table*}

%t3
\begin{table*}
\caption{Mean line-of-sight properties and derived quantities }
\label{table:Table3}
{
\small
\begin{tabular}{lccccccccc}
\hline
Sample & E(B-V) & N(HI)         & \NHIcl        & 2N(\HH) & \WCO & N(DNM) & \fH2 & \fDNM & \XCO\\
&         mag    &$10^{21}\pcc$ & $10^{21}\pcc$ &  $10^{21}\pcc$&K-\kms & $10^{21}\pcc$  & & & \\
\hline
Galactic anticenter & 0.362 & 1.324 & 1.119 & 0.798 & 5.185 & 0.268 & 0.38 & 0.13 & 1.93 \\
\#  &  33 &  33 &  33 &  33 &   9 &  33 &  33 &  33 &   9 \\
\hline
Chamaeleon & 0.271 & 1.282 & 0.619 & 0.393 & 2.400 & 0.205 & 0.23 & 0.12 & 2.30 \\
\#  &  13 &  13 &  13 &  13 &   1 &  13 &  13 &  13 &   1 \\
\hline
All LOS & 0.336 & 1.312 & 0.977 & 0.684 & 4.906 & 0.250 & 0.34 & 0.13 & 1.95 \\
\#  &  46 &  46 &  46 &  46 &  10 &  46 &  46 &  46 &  10 \\
\hline
\EBV\ $\le0.2$mag & 0.131 & 0.818 & 0.710 & 0.168 &   & 0.082 & 0.18 & 0.08 &  \\
\#  &  13 &  13 &  13 &  13 &    &  13 &  13 &  13 &    \\
\hline
Lacking \hcop $^1$ & 0.134 & 0.918 & 0.859 & $\leq$0.152 &   & 0.076 &  $\le$0.14 & $\ge$0.08 &  \\
\#  &   5 &   5 &   5 &   5 &    &   5 &   5 &   5 &    \\
\hline
DNM $\leq 10^{20}\pcc$ & 0.273 & 1.373 & 0.953 & 0.539 & 5.070 & 0.004 & 0.28 & 0.00 & 1.67 \\
\#  &  17 &  17 &  17 &  17 &   4 &  17 &  17 &  17 &   4 \\
\hline
CO detected earlier$^2$ & 0.502 & 1.369 & 0.928 & 1.910 & 4.906 & 0.204 & 0.58 & 0.06 & 1.95 \\
\#  &  10 &  10 &  10 &  10 &  10 &  10 &  10 &  10 &  10 \\
\hline
30m CO detected & 0.467 & 1.491 & 1.197 & 1.312 & 2.170 & 0.318 & 0.47 & 0.11 & 3.02 \\
\#  &  18 &  18 &  18 &  18 &  18 &  18 &  18 &  18 &  18 \\
\hline
30m \WCO\ $>$1K-\kms & 0.685 & 1.690 & 1.082 & 2.824 & 5.522 & 0.291 & 0.63  & 0.06 & 2.56 \\
\#  &   6 &   6 &   6 &   6 &   6 &   6 &   6 &   6 &   6 \\
\hline
30m \WCO\ $\leq$1K-\kms & 0.358 & 1.391 & 1.255 & 0.555 & 0.494 & 0.332 & 0.29 & 0.17 & 5.62 \\
\#  &  12 &  12 &  12 &  12 &  12 &  12 &  12 &  12 &  12 \\
\hline
30m UL $\leq$1K-\kms & 0.287 & 1.226 & 1.107 & 0.198 & $<$0.140 & 0.273 & 0.14 & 0.19 & $>$7.07 \\
\#  &  10 &  10 &  10 &  10 &  10 &  10 &  10 &  10 &  10 \\
\hline
Sample & E(B-V) & N(HI) & \NHIcl & 2N(\HH) & \WCO & N(DNM) & \fH2 & \fDNM & X$_{\rm CO}$\\
& mag&$10^{21}\pcc$ & $10^{21}\pcc$ &  $10^{21}\pcc$& & $10^{21}\pcc$  & & & \\
\hline
\end{tabular}
\\
$^1$ Limits in this table use $3\sigma$ upper limits on undetected quantities \\
$^2$ All sightlines detected earlier in CO had \WCO\ $>$1K-\kms \\ 
}
\end{table*}

%%%%%%%%%%%%%%%%%%%%%%%%%%%%
%%%%%%%%%%%%%%%

%s5
\section{Summary}
\label{section:Section5}
We took 89.2 GHz ALMA \hcop\ absorption spectra toward 33 compact millimeter-wave 
extragalactic continuum sources seen against the Galactic anticenter (Figure~\ref{fig:Figure1} 
and the tables above).  We also observed J=1-0 CO emission at the IRAM 30m telescope in 
the 28 anticenter directions where \hcop\ was detected. Inferring N(\HH) from 
N(\hcop) using the ratio N(\hcop)/N(\HH) $= 3\times 10^{-9}$, we combined these 
results with those from our earlier study of 13 directions where \hcop\ absorption 
was detected in the outskirts of Chamaeleon. We compared the inferred  N(\HH) with 
prior determinations of the column densities of the  dark neutral medium, the neutral gas 
of uncertain (atomic or molecular) character that had been found to be missing in 
HI and/or CO emission when compared with the  submillimeter dust and $\gamma$-ray 
emissivities of large-scale molecular cloud complexes.

Binning the HI, \HH,\ and DNM column densities in reddening, we showed in Figure 
3 that the mean DNM and molecular gas column densities were comparable and varied 
compatibly by factors of 50-100 over the observed range \EBV\ = 0.09 - 1 
mag, while N(H(I) varied by only factors of 2-3. The means of N(\HH) and N(DNM) 
are small at 
low mean reddening, and increase in similar fashion up to \EBV = 0.5 mag
where molecular gas begins to dominate and CO emission is strong. N(\HH) continues 
to increase with higher reddening, but N(DNM) and the column density of 
cloud-associated HI fall where \HH\ dominates (sequestering hydrogen) and CO emission
is stronger and more closely representative of N(\HH). 

We made detailed individual sightline-level comparisons of N(DNM) with \EBV, N(HI), 
and N(\HH) in Figures 4-6. Sightlines with appreciable DNM appear 
at all N(HI) (Figure~\ref{fig:Figure4}); the overall mean DNM fraction 
\fDNM\ = N(DNM)/(N(HI)+2N(\HH)) = 0.12 is modest.  Sightlines with appreciable DNM are
lacking when \EBV\ $\la$ 0.15 mag (Figure~\ref{fig:Figure5}) and when 2N(\HH) $\la 10^{20}\pcc$ 
or 2N(\HH) $\ga 10^{21}\pcc$ (Figure~\ref{fig:Figure6}). In Figure~\ref{fig:Figure6} we compared N(DNM) and N(\HH) 
in directions lacking detected CO emission in order to eliminate the case that 
\HH\ was already represented by CO emission in the DNM analysis.  This figure showed 
that there were 2-3 sightlines in each subsample (Chamaeleon and anticenter) or 5-6/46 
overall where 2N(\HH) $<\leq$ N(DNM)/2 and \HH\ accounted for the minority of the 
DNM. There are also a few directions at 2N(\HH) $\approx 10^{21}\pcc$ where 
2N(\HH) $>$ 2N(DNM).  Overall, the amounts of DNM and \HH\ are similar 
$<$2N(\HH)$>$ = 1.3 $<$N(DNM)$>$ for the unambiguous cases lacking CO emission.
The form of the DNM is overwhelmingly molecular hydrogen.

Directions with 
$<$\WCO$>$ $>$1 K-\kms\ have two times smaller mean DNM fractions $<$\fDNM$>$ = 
0.06, while sightlines with $<$\WCO$>$ $<$ 1 K-\kms\ have three times higher 
\fDNM\ $\ga$ 0.17-0.19 (Table~\ref{table:Table3}). The relatively few sightlines with \WCO\ 
$\ge$ 1 K-\kms\ contain two-thirds of the \HH\ detected in \hcop,\ and detecting 
90\% of the \HH\  would require detecting CO at levels \WCO\ 
$\approx$ 0.2-0.3 K-\kms . 

The CO-\HH\ conversion factor falls steadily with increasing \EBV\ or \WCO\ in 
Figure~\ref{fig:Figure7}. Because the \HH\ is concentrated in the sightlines with 
\WCO\ $\ge$ 1 K-\kms, the overall mean CO-\HH\ conversion factors in
our work are $<$N(\HH)$>$/$<$\WCO$>$ = $2-3 \times 10^{20}$\HH$\pcc$ for 
samples of sightlines with detectable CO emission.  These values are comparable 
to previously determined large-scale Galactic averages, and are substantially 
higher than the global values determined by the cloud-level analyses quoted 
here to derive N(DNM).  We ascribed these differences in part to the present 
sampling of widely scattered sightlines (i.e., that we did not do a cloud-level 
analysis) and perhaps to differences in the operational definition of the 
conversion factor, as discussed in Section~\ref{section:Section4.1}.  

Subsequent papers derived from the observations discussed here will focus on 
the physics and chemistry of the molecules observed in the course of this work, 
chiefly \hcop, \cch,\ and HCN.

%%%%%%%%%%%d%%%%%%%%%%%%%%% 
\begin{acknowledgements}

The National Radio Astronomy Observatory (NRAO) is a facility of the
National Science Foundation, operated by Associated Universities, Inc.

This work was supported by the French program “Physique et Chimie du Milieu 
Interstellaire” (PCMI) funded by the Conseil National de la Recherche Scientifique 
(CNRS) and Centre National d'Etudes Spatiales (CNES).

This work is based in part on observations carried out under project number 
003-19 with the IRAM 30m telescope]. IRAM is supported by INSU/CNRS (France), 
MPG (Germany) and IGN (Spain).

This paper makes use of the following ALMA data

\smallskip\noindent
-- ADS/JAO.ALMA\#2016.1.00714.S\\ 
-- ADS/JAO.ALMA\#2018.1.00115.S

\smallskip\noindent
ALMA is a partnership of ESO (representing its member states), NSF (USA)
and NINS (Japan), together with NRC (Canada), NSC and ASIAA (Taiwan), and
KASI (Republic of Korea), in cooperation with the Republic of Chile.  The
Joint ALMA Observatory is operated by ESO, AUI/NRAO and NAOJ.

We thank Isabelle Grenier for providing results of the DNM analysis 
and we thank the anonymous referee for many helpful remarks.
HSL is grateful for the hospitality of the ITU-R and the Hotel de la 
Cigogne in Geneva during the completion of this manuscript.

\end{acknowledgements}

%\bibliographystyle{apj}
%\bibliography{mnemonic,absorption}

\begin{appendix}

%Appendix A
\section{Statistics of HCO$^+$ detections}
\label{section:AppendixA}

\begin{figure}
\includegraphics[height=8cm]{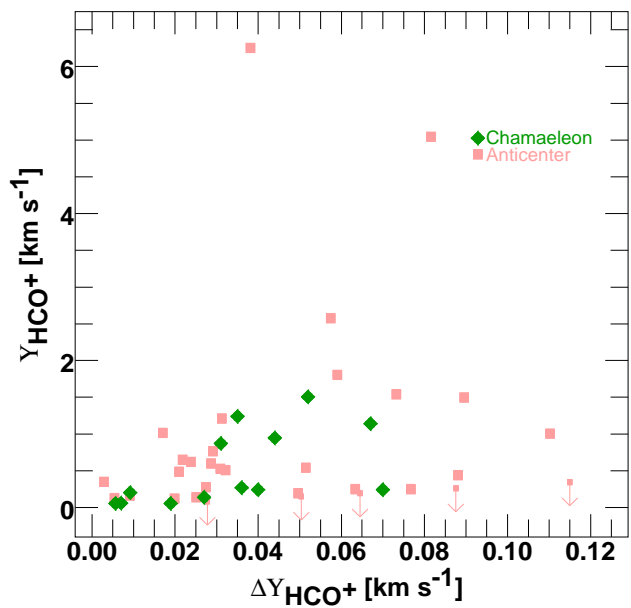}
 \caption[]{
Integrated \hcop\ optical depth \Uhcop\ plotted against its rms error. 
The upper limits along five sightlines not detected in \hcop\ are $3\sigma$.}
\label{fig:FigureA.1}
\end{figure}

\hcop\ absorption was not detected in five directions, all in the 
Galactic anticenter sample. Figure~\ref{fig:FigureA.1} 
shows \Uhcop\ plotted against its rms error
$\Delta$\Uhcop, in essence against the strength of the continuum background 
since all sightlines were observed for the same length of time. The plot 
shows that 
even the noisiest sightlines miss relatively small amounts
of molecular gas.  The bias in the plot, whereby $<$\Uhcop$>$ increases
with $\Delta$\Uhcop, was to some extent built into the observing 
to increase the yield.  That is, the source selection began with  
flux-limited samples that would achieve a minimum signal-to-noise ratio 
on the weakest source, and added 50\% additional targets at lower
flux \Snu\ and higher \EBV/\Snu\ that could be serendipitously 
observed without a proportionate increase in the required observing 
time.

%%AppendixB
\section{Spectra of HCO$^+$ and CO}
\label{section:AppendixB}

\begin{figure*}
\includegraphics[height=18cm]{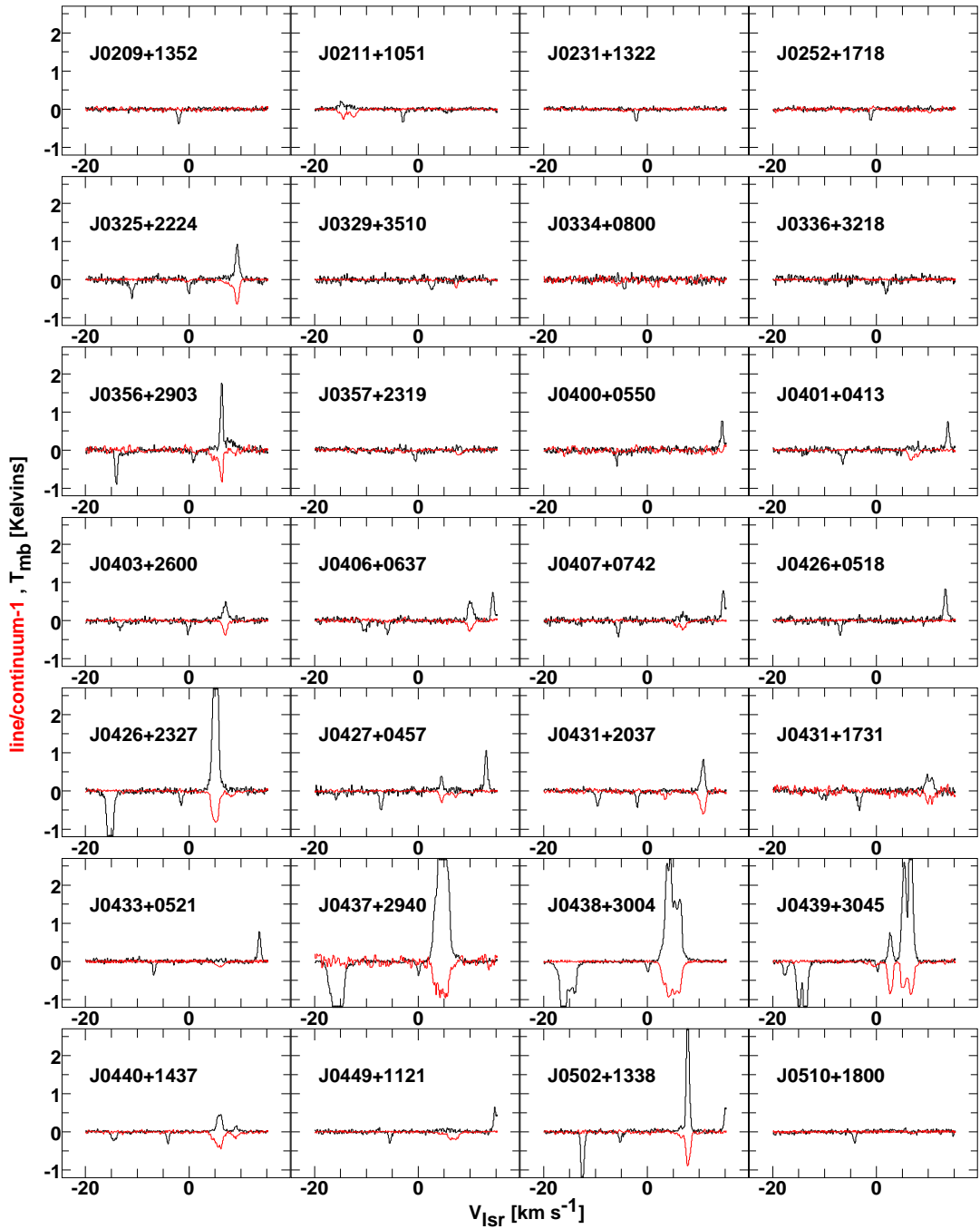}
 \caption[]{Spectra of  \hcop\ absorption (red) and CO emission (black) from the Galactic 
 anticenter sample. For an explanation of the appearance of the 
 frequency-switched IRAM 30m emission spectra, see  Appendix B.} %~\ref{section:AppendixZ}.}
\label{fig:FigureB.1}
\end{figure*}

Spectra of \hcop\ absorption and CO emission in the 28 directions with 
detected \hcop\ absorption are shown in Figure~\ref{fig:FigureB.1}. 
The \hcop\ absorption is the 
negative-going signal shown in red.  The black histogram shows the CO emission.
The frequency-switched IRAM 30m CO spectra were delivered only after folding in 
frequency, preventing a separation of the phases of the frequency-switching cycle 
using the methods of \cite{Lis97FS}. Negative-going features in the CO spectrum
are artifacts; only the positive-going CO signal coincident with \hcop\ 
absorption is interstellar.

%Appendix C
\section{Area used to derive large-scale statistics}
\label{section:AppendixC}

\begin{figure}
\includegraphics[height=7.2cm]{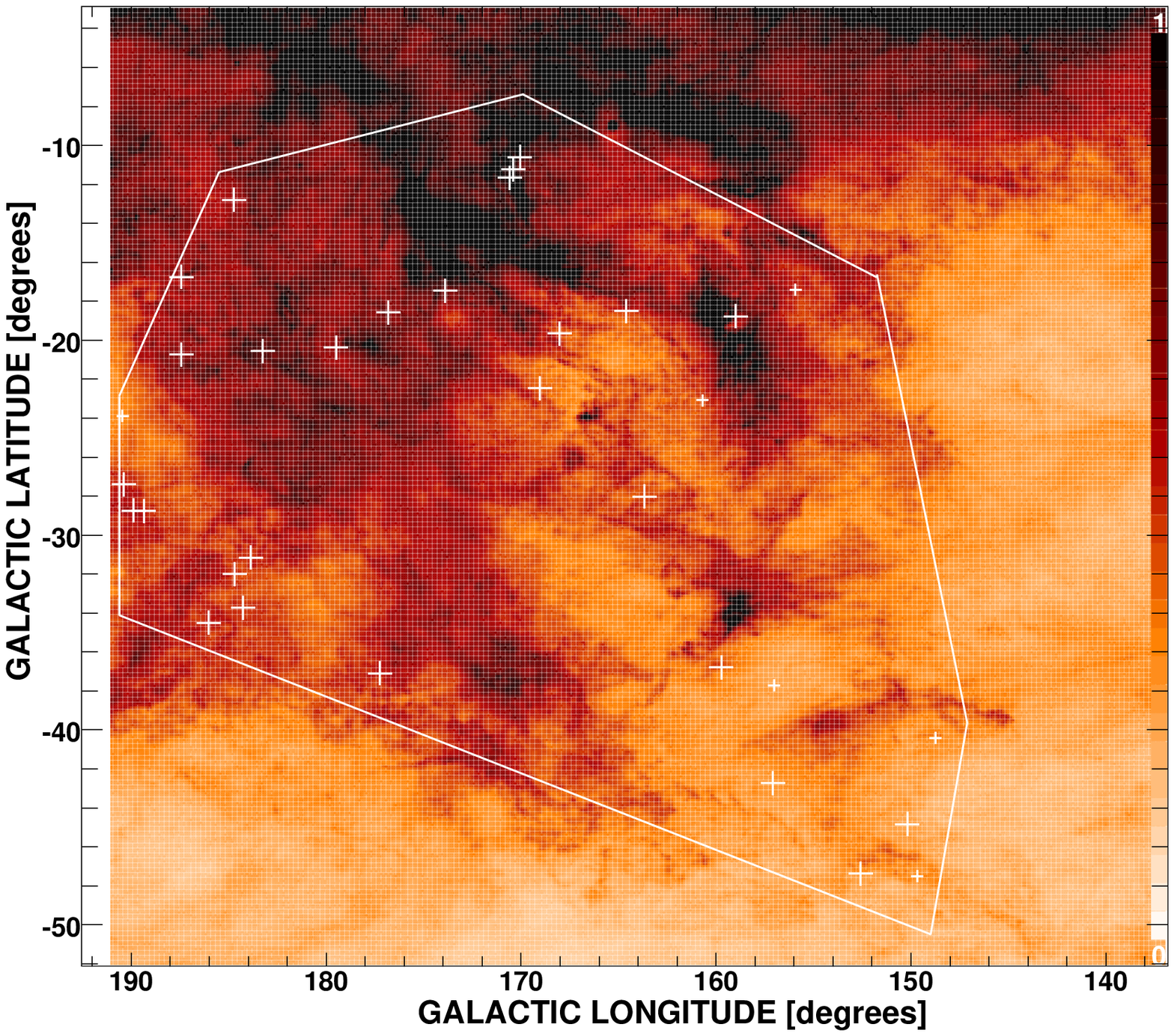}
 \caption[]{As in Figure~\ref{fig:Figure1},  left, but   showing the outlines of an area
containing the observed sightlines    used to derive large-scale 
properties of the anticenter region (see Section~\ref{section:Section4.3}).}
\label{fig:FigureC.1}
\end{figure}

To derive the statistical distributions shown 
in Figure~\ref{fig:Figure8} and discussed
in Section~\ref{section:Section4.4}, we drew a hull around the observed sightlines 
as illustrated in Figure~\ref{fig:FigureC.1}
and summed over the interior pixels.

\end{appendix}

\end{document}